\documentclass[twocolumn,amsmath,amssymb, nobibnotes, aps, prl, longbibliography,superscriptaddress]{revtex4-2}
\usepackage{graphicx}
\usepackage{lipsum}
\usepackage{amsmath}
\usepackage{xcolor}
\usepackage{bm}
\usepackage{color}

\renewcommand{\vec}[1]{\bm{#1}}

\newcommand\diff{\mathrm{d}}
\definecolor{myblue}{rgb}{0.125,0.125,0.75}
\usepackage{hyperref}
\hypersetup{colorlinks=true, linkcolor=myblue,urlcolor=myblue, citecolor=myblue}

\begin{document}
\title{Excluded Volume Induces Buckling in Optically Driven Colloidal Rings}
\author{Eric Cereceda-L\'opez}
\thanks{Both authors equally contributed to this work}
\affiliation{Departament de F\'{i}sica de la Mat\`{e}ria Condensada, Universitat de Barcelona, 08028 Spain}
\affiliation{Institut de Nanoci\`{e}ncia i Nanotecnologia, Universitat de Barcelona, Barcelona, Spain}
\author{Mattia Ostinato}
\thanks{Both authors equally contributed to this work}
\affiliation{Departament de F\'{i}sica de la Mat\`{e}ria Condensada, Universitat de Barcelona, 08028 Spain}
\affiliation{Universitat de Barcelona Institute of Complex Systems (UBICS), Universitat de Barcelona, Barcelona, Spain}
\author{Antonio Ortiz-Ambriz}
\affiliation{Departament de F\'{i}sica de la Mat\`{e}ria Condensada, Universitat de Barcelona, 08028 Spain}
\affiliation{Tecnologico de Monterrey, Escuela de Ingenier\'ia y Ciencias, Campus Monterrey, 64849 Mexico}
\author{Arthur V. Straube}
\email{straube@zib.de}
\affiliation{Zuse Institute Berlin, Takustra{\ss}e 7, 14195 Berlin, Germany}
\author{Matteo Palassini}
\affiliation{Departament de F\'{i}sica de la Mat\`{e}ria Condensada, Universitat de Barcelona, 08028 Spain}
\affiliation{Universitat de Barcelona Institute of Complex Systems (UBICS), Universitat de Barcelona, Barcelona, Spain}
\author{Pietro Tierno}
\email{ptierno@ub.edu}
\affiliation{Departament de F\'{i}sica de la Mat\`{e}ria Condensada, Universitat de Barcelona, 08028 Spain}
\affiliation{Universitat de Barcelona Institute of Complex Systems (UBICS), Universitat de Barcelona, Barcelona, Spain}
\affiliation{Institut de Nanoci\`{e}ncia i Nanotecnologia, Universitat de Barcelona, Barcelona, Spain}
\date{\today}
\begin{abstract}
%We study the out of equilibrium deformations in a shrinking ring of optically trapped, interacting colloidal particles. Steerable optical tweezers are used to confine dielectric microparticles along a circle of discrete harmonic potential wells, and to reduce the ring radius at a controlled quench speed. Excluded-volume interactions induce particle sliding from the equilibrium positions and zig-zag roughening of the colloidal structure. From the particle positions, we experimentally measure the ring distortion and roughening, and describe them by combining analytical theory with numerical simulations.  Our work unveils the underlying mechanism of interfacial deformation in driven  microscopic discrete rings.
In our combined experimental, theoretical and numerical work, we study the out of equilibrium deformations in a shrinking ring of optically trapped, interacting colloidal particles. Steerable optical tweezers are used to confine dielectric microparticles along a circle of discrete harmonic potential wells, and to reduce the ring radius at a controlled quench speed. We show that excluded-volume interactions are enough to induce particle sliding from their equilibrium positions and nonequilibrium zigzag roughening of the colloidal structure. Our work unveils the underlying mechanism of interfacial deformation in radially driven microscopic discrete rings.
\end{abstract}
\maketitle
\textbf{\textit{Introduction.-}} 
Surface roughening occurs in a wide variety of physical~\cite{Chui1978,Kardar1986} and biological systems~\cite{Vicsek1990,Wakita1997,Rapin2021}
and it is often caused by noise or external perturbations which deform 
an initially flat and spatially uniform interface~\cite{Vicsek1992,Barabasi1995}. 
Such process has received much attention on extended systems, i.e. when 
macroscopically large domains grow and deform,
including 
fluid~\cite{Flekkoy1995,Starr1996,Stelitano2000,Sagis2007}
and solid~\cite{Henk1977,Hoogeman1999,Bonzel2003,Balibar2005} interfaces,
grain boundaries~\cite{Gokhale2012,Lavergne2017,Liao2018} 
or active matter~\cite{Andac2018,Fausti2021,Lemma2022}.
However, finite-size systems made of expanding or shrinking circular domains also show 
kinetic roughening when driven out of equilibrium by internal fluctuations or pressure imbalance. Physical examples 
include unilamellar vesicles~\cite{Pecreaux2004,Faris2009}, confined bacteria colonies~\cite{Huergo2011,Giomi2019},
nematic
liquid crystals droplets~\cite{Takeuchi2010,Kazumasa2011}
or self-avoiding ring polymers~\cite{Witz2011,Santalla2014}.
In most of the cases, interfacial deformations are described with 
continuum models based on 
stochastic differential equations, such as the celebrated
Kardar, Parisi, and Zhang equation~\cite{Kardar1986}
or more sophisticated theories~\cite{Zhang1995,Krug2006}. 
An alternative, although less exploited approach, consists 
in considering the microscopic constituent of an interface, 
to understand how the global deformation arises directly  
from their pair interactions. Few works along this direction showed that spherical~\cite{Toussaint2004} or anisotropic~\cite{Yunker2013} particles can indeed act as 
microscopic model system for interfacial deformations. 

When two colloidal particles approach each other, they interact due to 
long-range electrostatic \cite{Russel1989} or hydrodynamic \cite{Lutz2006,Roichman2007,Sokolov2011,Nagar2014,Lips2022} forces or short-range steric effects arising from adsorbed polymer layers~\cite{Ekaterina1990}. 
However, when brought at very close contact, a strong repulsion emerges from the excluded volume, namely the mutual particle impenetrability. 
While such effect is negligible 
at extreme particle dilution, it strongly affects the 
diffusion~\cite{Dzubiella2003,Bruna2012,Rusciano2022},  transport~\cite{Kegel2000,Lips2018,Schmidt2021,AntonovPRL2022}, %melting~\cite{Pusey1986,Julio1995,Bernard2011,Thorneywork2017} 
and rheological~\cite{Mattsson2009,Chen2010} properties
of dense colloidal suspensions. 
Excluded volume creates geometrically frustrated buckled systems under strong confinement~\cite{Han2008,Zhou2017}, 
or can be used to engineer novel colloidal phases and structures~\cite{Sacanna2010,Feng2015}. 
%In the presence of a shear,  
%the particles may form 
%"force chain" at close contact leading to jamming~\cite{Cates1998}.
Further on, excluded-volume interactions go beyond the colloidal domain, being determinant in many granular~\cite{Rosato1987,Duran1993}, polymeric~\cite{Bolhuis2002,Pollak2014} or biological~\cite{Reisner2007,Tree2013} systems. 

Many studies on colloidal particles confined to a ring have focused on the particle displacement along the \emph{tangential} direction, keeping fixed the radius of the ring~\cite{Lutz2006, Roichman2007, Sokolov2011, Nagar2014, Lips2022}. Recently, this setup has shown striking effects as the emergence of propagating cluster defects \cite{AntonovPRL2022, AntonovNJP2022} when compared to other related discrete systems based on linear, nearest-neighbor coupling  \cite{Carpio2001}. In contrast, here we investigate the much less studied mechanism of out of equilibrium deformations induced by \emph{radially} shrinking a ring of discrete particles at different compression speeds. We use the synergy between the experiment, theory and numerical simulations to show that buckling predominantly arises from excluded volume interactions, without the need to consider long-range interactions, such as e.g. dipolar, electrostatic or hydrodynamic forces, unnecessary. Unlike roughening induced by thermal fluctuations in presence of dipolar interactions \cite{Toussaint2004}, the well-controlled quench speed becomes a key driving factor that determines driven buckling.

%\blue{Previous works on confining colloidal particles along an optical ring have focused on the particle
%displacement along the \emph{tangential} direction, keeping fixed
%the radius of the ring~\cite{Lutz2006,Roichman2007,Sokolov2011,Nagar2014,Lips2022}.
%This setup has recently shown striking distinction in defect propagation \cite{AntonovPRL2022, AntonovNJP2022}
%when compared to the similar system with long-range interactions \cite{Carpio2001}. In contrast, here we investigate
%the much less studied mechanism of deformation when the ring is compressed along the \emph{radial} direction.}
%Here we use a colloidal model system to investigate the out of equilibrium deformations induced by shrinking a ring of discrete %particles at different compression speeds. 
%We  use time shared optical tweezers to realize a series of quasistatic potential wells arranged along a circular ring with a tuneable %radius. The  controlled reduction of the ring radius 
%forces the particles to interact, and displaces them from their equilibrium positions at the onset of close contact. We use numerical %simulations to show that the dynamics are governed solely by excluded-volume interactions which emerge at close contact. Further, we %develop a theoretical model based on the microscopic particle displacements that predicts well the onset of 
%deformations induced by the non-equilibrium compression. 

%%%%%%%%%%%%%%%%%%%%%%%%%%%%%%%%
% Fig. 1
%%%%%%%%%%%%%%%%%%%%%%%%%%%%%%%
\begin{figure}[t]
\includegraphics[width=\columnwidth]{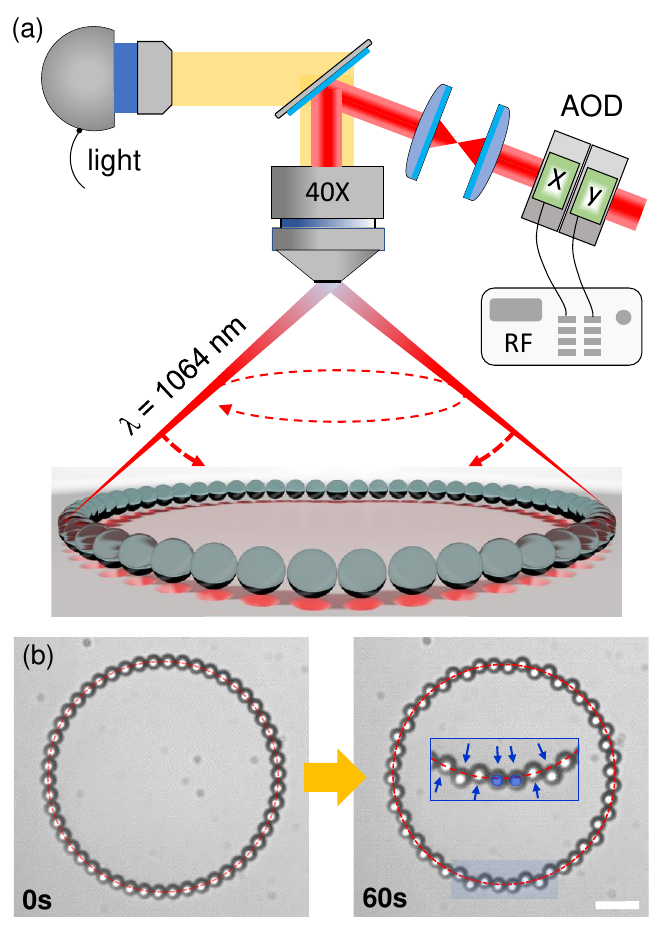}
\caption{Schematic of the experimental system: 
one infrared laser beam is rapidly 
steered via a two channel $(x,y)$ acousto-optic deflector (AOD)
controlled via a radio frequency generator (RF). The laser is scanned across equispaced points forming a ring of quasistatic double wells. 
(b) Two images showing the roughening of $N=50$ polystyrene particles when the optical ring is reduced from $R (0)= 34.7 \,\rm{\mu m}$ (left) to $R (\tau_q)= 31.0\, \rm{\mu} m$ (right) at a speed of $v_q=0.06 \, \rm{\mu m \, s^{-1}}$ (quench time $\tau_q=60 \,{\rm s}$). Small arrows in the inset of the right image indicate displaced particles, with one central defect highlighted. The dashed red line is the ring circumference,
see also VideoS1 in SM~\cite{EPAPS}.}
\label{figure1}
\end{figure}
%%%%%%%%%%%%%%%%%%%%%%%%%%%%%

\textbf{\textit{Experiments.-}} 
Our colloidal ring of radius $R$ is composed of $N=50$ polystyrene particles 
having $d=4.0 \, \rm{\mu m}$ diameter and dispersed in highly deionized water (MilliQ, Millipore). The particle solution is placed by capillarity inside a thin ($\sim 100 \rm{\mu m}$) chamber made of a borosilicate glass slide and a coverslip separated with a layer of parafilm. The experiments are performed at a room temperature of $T=293$K. We trap the particles using 
time shared optical tweezers which are created by 
passing an infrared laser beam (ManLight ML5-CW-P/TKS-
OTS operating at $3\, {\rm W}$, wavelength $\lambda = 1064 \,{\rm nm}$)
through a pair of acousto-optic deflectors (AODs, AA Optoelectronics DTSXY-400-
1064) both driven and synchronized with a radio-frequency (RF) generator (DDSPA2X-D431b-34).
The beam is then focused from above by a microscope objective (Nikon
$40 \times$ CFI APO), Fig.~\ref{figure1}(a). 
The bottom of the experimental chamber 
is observed with a second microscope objective (Nikon $40\times$ Plan Fluor) which projects an image onto a
CMOS Camera (Ximea MQ003MG-CM), in a custom-built inverted optical microscope. 
The laser tweezers visit $50$ equispaced 
positions along a ring,
spending $20 \, \rm{\mu s}$ in each point, such that 
each trap
is visited once every $\sim 1$ms.
Thus, the potential wells can be considered as quasistatic 
since the beam scanning is much faster than the 
typical self-diffusion time of the particles,
$\tau_D = d^2/(4D_{\rm{eff}}) \sim 30 \, {\rm s}$, estimated from the effective particle diffusion coefficient $D_{\rm eff} \simeq 0.13 \,\rm{\mu m^2 \, s^{-1}}$~\cite{Lips2022}. 
%The latter quantity was previously measured from the mean square displacement of single particle performing free diffusion~\cite{Lips2022}. As a consequence, each particle is stably confined at a given position along the ring, i.e. it is trapped individually in an harmonic-like well.

\begin{figure}[tp]
\includegraphics[width=\columnwidth]{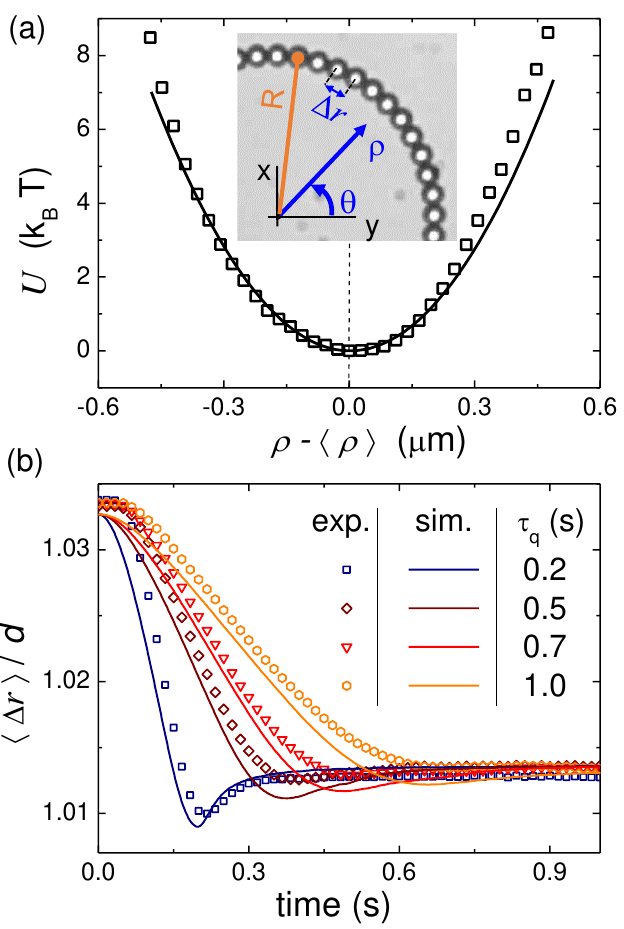}
\caption{(a) Confining potential $U(\rho)$ along the radial ($\rho$) direction  
calculated from the particle 
fluctuations before compression, here $\langle R \rangle=34.7 \rm{\mu m}$. 
Scattered squares
are experimental data, the continuous line is a non linear regression calculating $U(\rho)$ from Eq.~\eqref{Rayleigh}. 
The inset shows a colloidal ring with highlighted 
cartesian ($x,y$) and polar ($\rho,\theta$) coordinate systems, being $\Delta r$ the nearest neighbor distance between the centers of the particles along the ring.
(b) Average distance between the centers of
nearest neighbor particles $ \langle \Delta r \rangle$ 
versus time for four different quench times $\tau_q$ from experiments (symbols) and 
simulations (lines).}
\label{figure2}
\end{figure}

\textbf{\textit{Ring deformation.-}}  We start by analyzing the optical potential confining the colloidal particles
within a ring of mean radius 
$34.7 \, \rm{\mu m}$.
As shown in the inset in Fig.~\ref{figure2}(a), we work in polar coordinates with the origin at the center of the ring. Before compression, we determine the spring constant $\kappa$ of the optical potential confining each particle by monitoring the equilibrium, radial particle fluctuations across the ring.
%Assuming equilibrium conditions, 
One can obtain the confining potential $U(\vec{r})$ by measuring the stationary probability distribution directly from the particle trajectories~\cite{Crocker1994}. 
To adapt such procedure to the radial configuration, we formulate the overdamped
%Langevin
equation of motion for an individual Brownian particle with the position $\bm{r}=(x,y)$ captured in a radial harmonic trap centered at $\vec{R}$ (with $|\bm{R}|=R$):
% \equiv R (\cos{(2 \pi k/N)},  \sin{(2 \pi k  /N )})$ ($k=1,\dots,N$),
%and stiffness $\kappa$, 
$\gamma \dot{\vec{r}}=-\kappa (\vec{r}-\vec{R})+ \vec{\xi}$, 
where $\kappa$ is the trap stiffness, $\gamma = 3\pi \eta d$ the drag coefficient 
of the particle, $\eta$ the viscosity of water, 
%$\vec{R}=R \unit{\rho}$ with $\unit{\rho} =\vec{r}/r$ 
and $\vec{\xi}$ a stochastic force with zero average and delta correlated. 
%due to thermal fluctuations such that, 
%$\langle \vec{\xi}(t) \rangle =0$ and $\langle \vec{\xi}(t) \otimes \vec{\xi}(t')  \rangle = 2 D \vec{I} \delta(t-t')$ with $D= \kB T/\gamma$ the free diffusion coefficient
%and $\vec{I}$ the identity matrix.
The corresponding stationary distribution of the particle position 
is $P(\bm{r}) \propto \exp[-U(\vec{r})/k_B T]$, where
$U(\vec{r})= \kappa (\vec{r}-\vec{R})^2/2$. Passing to polar coordinates for 
$(\rho=|\vec{r}|,\theta)$ and making use of the radial symmetry of $P(\vec{r})$, we integrate
over $\theta$ to find
that the stationary distribution of the radial displacement 
has the ``Rayleigh'' form:
%By passing to the corresponding Fokker-Planck description in polar coordinates and finding its stationary, angle-independent solution, we 
%arrive at the radial ``Rayleigh distribution'' for the radial position of
%the particles 
\begin{equation}
P(\rho)= C \rho \,e^{-U(\rho)/k_B T}\, , \quad U(\rho)=\frac{\kappa}{2} (\rho-R)^2\,  
\label{Rayleigh}
\end{equation}
with $C$ being the normalization constant. For details,
see Supplemental Material in~\cite{EPAPS}.  
We invert Eq.~\eqref{Rayleigh} as:
$U(\rho)= -k_{B}T \ln{[R P(\rho)/(\rho P(R))]}$ and calculate  the radial potential
from the experimentally determined displacement distribution.  
%We confirm this trapping potential by calculating $U_{\rho}$
%from the experimentally determined displacement distribution.
Fig.~\ref{figure2}(a) shows the experimental data 
(open squares) with a non-linear regression using 
the expression in Eq.~\eqref{Rayleigh} for $U(\rho)$. From these data we
extract the optical spring constant $\kappa= (2.51 \pm 0.02) \cdot 10^{-4} \,\rm{pN \, nm^{-1}}$.

%%%%%%%%%%%%%%%%%%%%%%%%%%%%%%%% 
% Fig. 2
%%%%%%%%%%%%%%%%%%%%%%%%%%%%%%%
%%%%%%%%%%%%%%%%%%%%%%%%%%%%%

In a typical shrinking experiment, Fig.~\ref{figure1}(b)
we first equilibrate the trapped particles along a ring with an initial radius $R(t=0)=34.7 \, \rm{\mu m}$.
After that, we perform $50$ measuring cycles by repeatedly
decreasing the radius to $R (t=\tau_q)=31.0 \, \rm{\mu m}$ at a controlled quench speed $v_q=[R(0)-R(\tau_q)]/\tau_q$ 
and, after a short equilibration period, increasing it back to its initial position. 
The experiments run over $24$h during which we  
change  the quench time $\tau_q \in [0.2,10] \,{\rm s}$
and speed $v_q \in [0.37,18.5]\, \rm{\mu m \,s^{-1}}$
ensuring the same statistical averages for each value of $\tau_q$.

At $t=0$, the particles lay almost along their mean elevation $\langle \rho_i \rangle \approx R$, with 
a small thermal roughening $W(t=0)=W_0$ induced by thermal fluctuations, 
where we define the roughening $W(t)$ as
\begin{equation}
W^2(t) = \frac{1}{N} \sum_i \langle h^2_i(t) \rangle \, ,
\label{roughness}
\end{equation}
in which $h_i=\rho_i -\langle \rho_i\rangle$ is the radial displacement of particle $i$ ($=1,\dots,N$) from the mean. 

Upon decreasing the ring radius, the colloids start interacting and excluded volume displaces the particles either outside or inside the ring. For the chosen number of particles, the ground state of the system would be a perfect zig-zag chain with no defects~\cite{Straube2013}, which could be obtained for an adiabatically slow compression, $\tau_q \to \infty$.  
However, at a finite quench time defects
in the roughening emerge in the form of two or more particles displaced together, inset in Fig.~\ref{figure1}(b).  The effect of excluded-volume interactions can already be appreciated by measuring the evolution of the average distance $\langle \Delta r \rangle$  between the centers of neighboring particles, Fig.~\ref{figure2}(b). For the fastest compression occurring at $\tau_q=0.2 \,{\rm s}$, $\langle \Delta r \rangle$ displays a pronounced minimum close to the excluded-volume limit, $\langle \Delta r \rangle =d$ indicating a strong repulsive force around $t \sim \tau_q$. After that,   
the particles reach a steady distance of  $\langle \Delta r \rangle \sim 1.013 d$, regardless of the compression time. Reducing the compression speed 
gradually eliminates this minimum since the slower approach allow  the colloids to rearrange sliding out 
from their central position in the harmonic wells.

%%%%%%%%%%%%%%%%%%%%%%%%%%%%%%%%
% Fig. 3
%%%%%%%%%%%%%%%%%%%%%%%%%%%%%%%
\begin{figure*}[t]
\includegraphics[width=\textwidth]{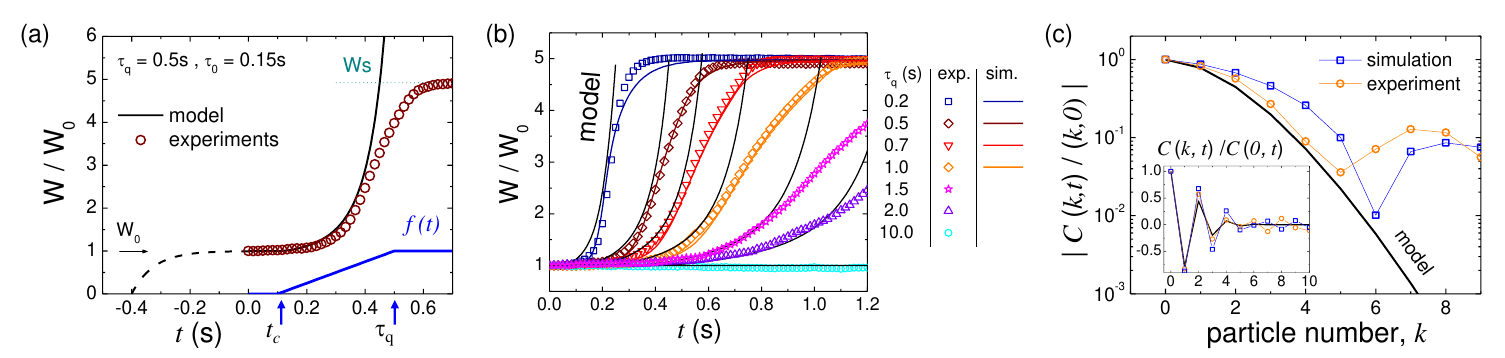}
\caption{(a) Normalized roughening, $W/W_0$ 
versus $t$ from experiments (symbols) and Eq.~\ref{exact} (black line) for $\tau_q=0.5$s
and $t_{eq}= 0.4$ s. The linear schedule $f(t)$ is shown in blue.
(b) $W/W_0$ \textit{vs} $t$ from experiments (symbols), 
simulations (lines in colour) and model (black line) for different $\tau_{q}$.
(c) Absolute value of the normalized, equal time correlation function, $C(k,t)/C(0,t)$ 
versus particle number $k$ in semilog scale. Empty symbols are experimental (circles) and simulation (squares) data for $\tau_{q}=0.5$ s, while continuous line is the numerical integration of Eq.~\eqref{exact2} at time $t=0.45$s at which the predicted $W(t)$ reaches $W_s$.
Bottom inset shows the correlations in normal scale.}
\label{figure3}
\end{figure*}
%%%%%%%%%%%%%%%%%%%%%%%%%%%%%%%

\textbf{\textit{Numerical Simulations.-}} 
To understand whether
the induced buckling can be described by
only taking into account excluded-volume interactions, we perform
Brownian dynamics simulations with the same parameters
as in the experiments. We extend our single-particle description to many interacting particles, in which each particle $i$ obeys the Langevin equation:
\begin{equation}
\gamma \bm{r}_i =- \kappa (\bm{r}_i - \bm{R}_i) +\bm{f}_{i,i+1}+\bm{f}_{i,i-1}+\bm{\xi}_i\,, \label{eq:LEs-full}
\end{equation}
where the force $\bm{f}_{i,j}=-U'_\textrm{HC}(r_{ij}) \bm{r}_{ij}/r_{ij}$ represents the hard-core repulsive interaction between nearest neighbors. To achieve the best mapping with the experimental data, we generalize the Weeks-Chandler-Andersen %(WCA)
potential~\cite{Weeks1971} as, $U_\textrm{HC}(r)=4\epsilon[(d/r)^{2q}-(d/r)^{q}+1/4]$ for $r\leq r_0=2^{1/q}d$ and zero otherwise. Here $\epsilon$ is the repulsion strength and $q$ the nonlinearity index. 
As shown in Fig.~\ref{figure2}(b) and \ref{figure3}(b), we find quantitative agreement between simulations and experiments by using $q=42$ and $\epsilon/\gamma = 2.5 \, {\rm \mu} {\rm m}^2 {\rm s}^{-1}$ as fitting parameters \cite{EPAPS}. This matching highlight that the particles experience an effective short-range repulsive potential, a quantity frequently subject to uncertainty due to the presence of residual particle charges and long-range interactions. 
Further, by varying the number $N$ of particles at constant $t_q$, the roughening shows similar crossover from the same thermal plateau $W_0$ at small times to the saturation limit $W_\text{S}$ at long times. While $W_\text{S}$ grows only slightly for larger $N$, it exhibits a stronger nonuniform decrease for smaller $N$, due to the strong, nonlinear nature of the short-range interactions.
%More details on the simulations are given in SM \cite{EPAPS}.
%\begin{equation}
%U(r)=\frac{576 \epsilon}{r_0^2} (r-r_0)^2 \, \, \, , \, \, \textrm{for}  \, \, \, r\leq r_0 \, º, \, \, \, \textrm{(otherwise 0)}
%\label{WCA}
%\end{equation}
%with cutoff distance $r_0$  and being $\epsilon$ the strength of the WCA 
%repulsive potential, see~\cite{EPAPS} for more details. 
%Fig.~\ref{figure2}(b) shows that 
%the numerical simulations are able to capture well the experimental results for all the tested $\tau_q$,
%excluding effects due to hydrodynamic or long range interactions
%as electrostatic one.

\textbf{\textit{Theory.-}} 
We introduce the following one-dimensional model 
for the radial displacements:
%of particles from
%their mean position, $h_i=\rho_i-\langle \rho_i \rangle$, is:
\begin{equation}
\gamma \dot{h}_i =   -\kappa g(t)(h_{i+1}-2h_i+h_{i-1}) -\kappa h_i+\xi_i\,.
\label{eq:h}
\end{equation}
%where $\gamma = 3\pi \eta d$ is the drag coefficient 
%of the particle, $\eta$ the viscosity of water, $\kappa$ the trap and $\xi_j(t)$ 
%a random force such that $\langle \xi_j(t) \rangle =0$
%and $\langle \xi_i(0) \xi_j(t) \rangle = 2\gamma k_B T \delta(t) \delta_{ij}$.
%equation of motion used in the simulations, Eq. \eqref{eq:LEs-full} onto the radial direction. 
Unlike a previous work on thermal induced roughening
of a linear chain [cf. Eq.~(2) in Ref.~\cite{Toussaint2004}], our Eq.~\eqref{eq:h} goes beyond the
nondriven framework, as it
%further
includes a time-dependent factor $g(t)$ that encapsulates
the increase in the effective interactions as the ring is compressed. Moreover the particle interaction considered here are repulsive ($g(t)>0$) in contrast to Ref.~\cite{Toussaint2004}. 

Equation~\eqref{eq:h} can be justified by projecting Eq. \eqref{eq:LEs-full} of the simulations onto the radial direction (see SM \cite{EPAPS}).
In this way we obtain $g(t)=0$ for $t<t_c$ and $g(t)=2 q^2 \epsilon (R_c/R(t) - 1) / (r_0^2 \kappa)$ for 
$t>t_c$, where $R(t)=R(0)-v_q t$ is the ring radius at time $t$, and $t_c$ is the time at which $R(t)=R_c \equiv N r_0/2 \pi$
and the particles start to interact.
Note that in Eq.~\eqref{eq:h} we have omitted a term $D/(h_i+R)$, which is negligible for our experimental conditions \cite{EPAPS}.

Using the discrete Fourier transform, from Eq.~\eqref{eq:h} we calculate the equal-time height-height correlation function $C(k,t)=N^{-1}\sum_i \langle h_i(t) h_{i+k}(t) \rangle$ of the ring:
\begin{equation}
C(k,t)=2 (-1)^k D \int_{-t_{eq}}^t  F_k(t,t') \, \diff t'\,
\label{exact2}
\end{equation} 
with $F_k(t,t')=I_k[ (4/\tau_0) \int_{t'}^t g(s) ds ] {\rm{exp}}[-2  (t-t')/\tau_0 +
(4/\tau_0) \int_{t'}^t g(s) ds ]$.
Here $\tau_0=\gamma/\kappa$ is the microscopic relaxation time, $I_k(x)$ is the modified Bessel function of the first kind and $t_{eq}$ is an equilibration time
which we suppose much larger than $\tau_0$ so that at $t=0$ the system is in thermal equilibrium.
For $k=0$, we obtain the roughening, cf. Eq.~\eqref{roughness}:
\begin{equation}
W^2(t)=\frac{2 W_0^2}{\tau_0}
\int_{-t_{eq}}^t F_0(t,t')\,dt' \, \, \, .
\label{exact}
\end{equation} 
In Eq.~\eqref{exact}, we fix from the experimental data $\tau_0=0.15 \, {\rm s}$ and $W_0=\sqrt{D_\text{eff} \tau_0} = 0.127 \, {\rm \mu m}$ which denotes the ``thermal'' roughening.

This equation predicts that, starting from $t=-t_{eq}$, $W(t)$ will first 
grow as $W(t)=W_0[1-\exp{(-2(t+t_{eq})/\tau_0)}]$, 
as shown by the dashed line in Fig.~\ref{figure3}(a), reach a thermal plateau $W_0$  for $-t_{eq}+\tau_0 \ll t \ll t_c$,
%
%shown by the dashed line in Fig.~\ref{figure3}(a) (analytical solution for $t<0$),
and grow again for $t>t_c$ due to the repulsive interaction.
For simplicity, we use a linear schedule  $c f(t)$ that approximates well the full $g(t)$ \cite{EPAPS},
where $f(t)=0$ for $t<t_c$, $(t-t_c)/(\tau_q - t_c)$ for $t_c\leq t \leq \tau_q$, and $1$ for $t>\tau_q$. This schedule is shown by the blue line in Fig.~\ref{figure3}(a). Here $c$ is a dimensionless constant which can be related to the pair interactions between the colloidal particles as $c=  (1-R(\tau_q)/R_c) 2 q^2 \epsilon/(r_0^2 \kappa)$.

Figure~\ref{figure3}(b) shows the roughening from rings 
compressed at different $\tau_q \in [0.2,10]$s.
We use Eq.~\eqref{exact} to fit the experimental data, the only adjustable parameters being $t_c/\tau_q$ and $c$.
We obtain a good quantitative description of the initial increase of $W(t)/W_0$ at different $\tau_q$ 
by using $t_c/\tau_q = 0.2$ and $c=0.9$, which determine 
$q=10.7$ and $\epsilon/\gamma = 5.4 {\rm \mu} {\rm m}^2 {\rm s}^{-1}$, respectively. 
Note that both values differ only slightly from those for the full numerical model, Eq. \eqref{eq:LEs-full}, despite a number of simplifications made to derive Eq.~\eqref{eq:h}. Interestingly, for $c<1/4$ the model is able to predict that $W(t)$ reaches an intrinsic saturation value $W_\infty = W_0(1-4 c)^{-1/4}$
for $t \gg \tau_q$, whereas for $c>1/4$ it grows without bounds, indicating that
the particles escape from the optical traps. Because the linear approximation for the short-range repulsion force in Eq.~\eqref{eq:h} is insufficient to describe the final stages when the interaction becomes very large, but quantitative agreement with the saturation value $W_s$ set by  the  experimental protocol can be achieved by simulations.
Note that it is not possible to measure experimentally
the initial growth until $W_0$ because the initial configuration
in the experiment is not perfectly flat.

Finally, we also use the model to contrast the experiments and simulations in terms of the equal time correlation function $C(k,t)$, Fig.~\ref{figure3}(c).
For a fixed $t$, both experiments and  simulations, display anticorrelations in $C(k,t)$, as described by  Eq.~\ref{exact2}, see the inset of Fig.~\ref{figure3}(c), which is a signature of the underlying zigzag pattern. 
The model fits rather well the experimentally
determined correlation function up to distance $k = 5$
particles. Above this range, the data start deviating and
both, simulations and experiments display a series of 
oscillations that may be due to correlated lateral displace-
ment of the particles, which are completely discarded by
our theory.

\textbf{\textit{Conclusions.-}} 
We have used a colloidal model system 
to investigate the deformations emerging from a collapsing chain of optically driven microscopic particles.
%We find that  the shrinking dynamics can be described by considering the sole excluded-volume interactions that  induce buckling thus, without any long-range electrostatic, steric (polymer mediated) or hydrodynamic effects. Our experimental results are supported by both theory and numerical situations. 
Via a tight combination of the experiments, theory, and simulations, we find that the nonequilibrium buckling is captured quantitatively well by considering the sole excluded-volume interactions, thus without any long-range dipolar, electrostatic, steric (polymer mediated) or hydrodynamic effects.
Different works have investigated the effect of topology on the collapse of a dried dense colloidal suspension  
\cite{Tsapis2005,Meng2014,Munglani2019,Wang2019}.
Here, we have focused on the mechanism leading to buckling starting from the interparticle interactions.
We expect that our general description of a finite shrinking 
ring at the discrete, i.e. single particle level, may be 
important to related research fields where 
finite droplets reduce and deform due to external or internal (i.e. active) pressure fields. 
%Indeed the shrinkage of small droplets due to evaporation is an important industrial problem~\cite{Lohse2015}, occurring in 
%different applications such as ink-jet printing~\cite{Gans2004,Gans20042}, spray technology~\cite{Hinds1982,Cho2008,Dubitsky2023} and DNA chip manufacturing~\cite{Dugas2005,Wohrle2020}. Our work may also be extended to biological systems,
%for example the deformation of circular colonies of bacteria or biological cells when excluded-volume interactions are dominant. 

We acknowledge stimulating discussions with Thomas M. Fischer, Hartmut L\"owen and Adolfo del Campo.  This project has received funding from the 
European Research Council (ERC) under the European Union's Horizon 2020 research and innovation programme (grant agreement no. 811234).
A.V.S. acknowledges support by the Deutsche Forschungsgemeinschaft (DFG) under Germany’s Excellence Strategy–MATH+: The Berlin Mathematics Research Center (EXC-2046/1)–Project No. 390685689 (Subproject AA1-18). 
M. P. acknowledges support by Ag\`encia de Gesti\'o d'Ajuts Universitaris i de
Recerca (AGAUR) (no. 2021 SGR 00247)
and Agencia Estatal de Investigaci\'on (MINECO) (no. PID2022-139913NB-I00).
P. T. acknowledge support the Generalitat de Catalunya under Program ``ICREA Acad\`emia'' and  AGAUR (no. 2021 SGR 00450).

\end{document}